# Humidity effects on tip-induced polarization switching in lithium niobate

A.V. Ievlev[1,2], A.N. Morozovska[3], V.Ya. Shur[2], S.V. Kalinin[1]

[1] The Center for Nanophase Materials Sciences, Oak Ridge National Laboratory, Oak Ridge, TN 37922

[2] Ferroelectric Laboratory, Institute of Natural Sciences, Ural Federal University, 51, Lenin Ave., 620000 Ekaterinburg, Russia

[3] Institute of Physics, National Academy of Sciences of Ukraine, 46, pr. Nauki, 03028 Kiev, Ukraine

Interest to ferroelectric materials has been increased significantly in last decades due to development of new generation of nonlinear optical and data storage devices. Scanning probe microscopy (SPM) can be used both for study of domain structures with nanoscale spatial resolution and for writing the isolated nanodomains by local application of the electric field. Tip-induced switching in the ambient still needs experimental investigations and theoretical explorations. Here we studied influence of the value of relative humidity in the SPM chamber on the process of tip-induced polarization switching. This phenomenon was attributed to existing of the water meniscus between tip and the sample surface in humid atmosphere. Presented results are important for further complex investigations of the ferroelectric materials and their applications.

Since its invention two decades ago[1–8] Piezoresponse Force Microscopy (PFM) has emerged as a powerful tool for experimental investigations of ferroelectric materials[6,9]. In the imaging mode (PFM) allows to visualize static domain structures with nanometer spatial resolution[6,10]. Application of sufficiently large voltage through conductive SPM tip can induce local polarization switching, and can be extended for creation of the tailored domain structures and ferroelectric data storage[11–16]. Finally, acquisition of the piezoresponse signals during polarization reversal allows to measure local hysteresis loops, which can be used for characterization of the switching process in the nanoscale area in vicinity of the tip[17,18].

The broad application of PFM for probing of the domain structures and polarization reversal in ferroelectrics demands deep understanding of the basic mechanisms involved. In particular, the key role of polarization screening on the PFM tip-induced polarization reversal has been demonstrated[19]. This analysis directly follows from the well-recognized role of the screening of depolarization field on stability of ferroelectric domain structure in general. While early works on ferroelectrics postulated the bulk screening by the charge carriers or defect dipole redistribution in the bulk, it has recently been realized that in ambient environment the external screening is realized by the ionic species[20]. In fact, polarization switching process can be represented as[21]:

$$([+P] - OH^-) + H_2O + 2e^- = ([-P] - H^+) + 2OH^- \qquad (1)$$

Here, $([+P] - OH^-)$ is the positive polarization charge bound with the screening hydroxyl group (see Refs. [22,23] for discussion of equilibrium degree of screening), and $([-P] - H^+)$ is the negative polarization charge bound with a screening proton. Note that maintaining local quasi-electroneutrality during polarization switching requires that the switching phenomena be coupled to surface electrochemical processes.

The characteristic aspect of such process is its non-locality, since the domain formed under the tip has a finite size. Hence, accommodation of free screening charges (i.e. hydroxyls in eq. 1) requires the lateral transport of ions across the surface, absorption by the tip or ionic emission from the surface. Correspondingly, tip-induced polarization switching can be strongly affected by the mobility of the charged ions on the sample surface. Recently, the role of surface treatment[24] and humidity in SPM chamber[25–27] on the polarization reversal process has been explored. Observed phenomena has been attributed to redistribution n of the switching electric field due to existing of conductive adsorption surface layer[24,27] and water meniscus in vicinity of SPM tip[25,26]. Here, we systematically explore the role of humidity on polarization switching in ferroelectrics. Obtained results at the first glance are opposite to the behavior revealed earlier; in our case humidity increasing leads to the hamper of the switching, while in[25] it supports switching and leads to formation of larger domains.

Here, we use single-crystals of the congruent lithium niobate $LiNbO_3$ (CLN) as a model uniaxial ferroelectric. The thickness of the sample has been decreased by thorough mechanical polishing to 15 μm. Experiments were performed by commercial scanning probe microscope (Bruker Nanoman, USA) using Multi-75G-E SPM tips (Budget Sensors, USA) with conductive platinum coating and radius of tip curvature $R_{tip} < 25$ nm. Local polarization reversal was carried out by triangular bipolar pulses with amplitude $U_{sw}$ ranged from 20 V to 100 V and duration $t_{sw} = 250$ ms. This shape of the switching pulse allowed carrying out simultaneous measurements of the local hysteresis loops in the band excitation PFM mode[28]. Experiments were carried out at room temperature and at relative humidity ranged from 0 to 90%. The variable humidity measurements were performed using General Electric MG110 hygrometer calibration tool for varying of the humidity of the nitrogen in the SPM chamber.

Local hysteresis loop measurements suggested three typical types of the switching processes (Fig. 1). The switching has been observed in the areas of the sample with spontaneous polarization directed downward ($Z^-$ polar surface) only. Two different switching behaviors have been observed: 1) "transient switching" with polarization reversal in the very beginning of the switching cycle (Fig. 1b) was observed in the dry atmosphere ($H \sim 0\%$) and was leading to formation of the isolated domains; 2) "normal switching" with conventional shape of the hysteresis loop has been observed at all others values of the $H$ (Fig. 1a). The switching on the Z+ polar surface hasn't been observed (Fig. 1c).

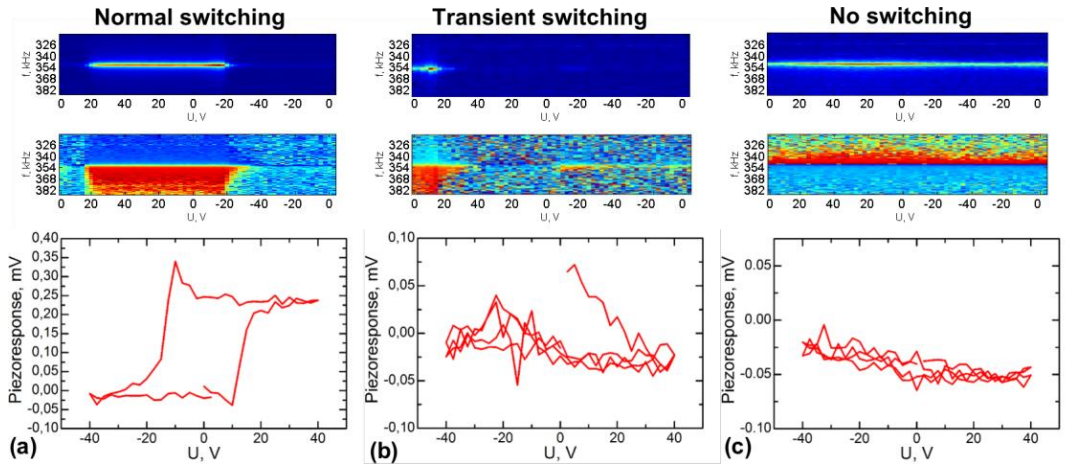

**Figure 1.** Amplitude and phase spectrograms and local hysteresis loops. Switching on $Z^-$ polar surface at (a) $H > 20\%$ and (b) $H = 0\%$. (c) Switching on $Z^+$ polar surface.

To explore the environmental effects on polarization reversal, chains of domains were written with same switching pulses parameters (Fig 2a). Size of the formed isolated domains with measured local hysteresis loop was used for characterizing of the switching process. Size averaging over few domains was used to enhance quality of the data. Distance between neighboring switching points was above 500 nm to avoid interaction between domains in the chain[21].

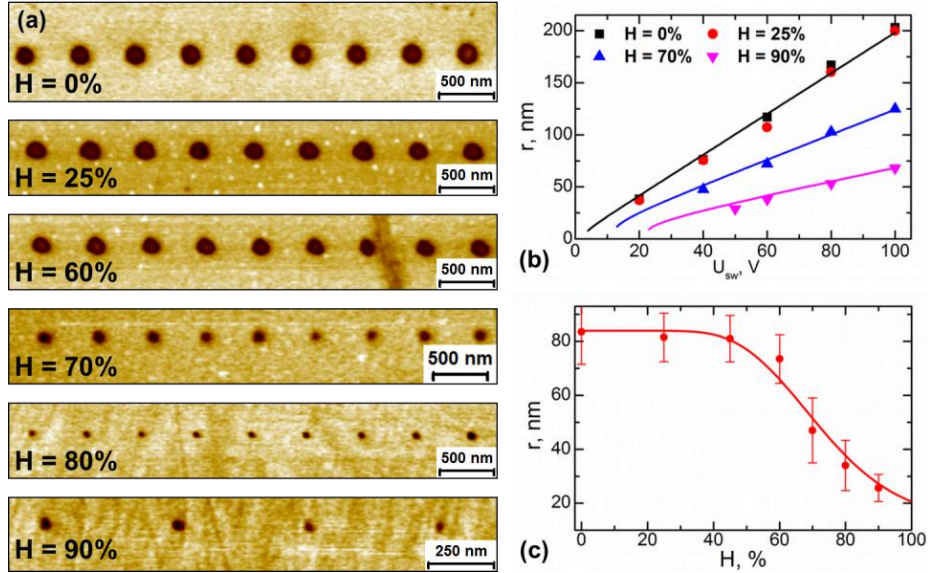

**Figure 2**. (a) Domains formed as a result of local polarization reversal at different humidities ($U_{sw} = 40$ V). Experimentally measured domain radius vs (b) bias and (c) relative humidity, fitted by eq (6).

Dependence of the domain radius on the amplitude (Fig. 2b) of switching pulse was found to follow well-known linear law[15,24]. The corresponding slope decreased with relative humidity. The critical voltage (cross point of the extrapolated $r(U_{sw})$ function and $r = 0$ axis) ranged from 3V in dry conditions ($H = 0 – 50\%$) to 27 V for $H=90\%$. Radius of the minimal stable domain was about 20 - 40 nm for all values of the relative humidity.

To get further insight into this behavior, the dependence of the domain radius on humidity in the SPM chamber for a fixed bias has been studied, as shown in Fig. 2a,c. Switching in dry air led to formation of the domains with radius about 80 nm. Increasing of the humidity up to 60% didn't reveal any essential changes in the domains radii. Further increasing of the humidity led to essential decreasing of the domain size and practically disappearing of the domains at 90% (Fig. 2a).

Additional information about local polarization reversal has been obtained by meaning of the local hysteresis loops measured at different values of relative humidity *H* (Fig. 3a-c). Mathematical analysis of the loops[29] allowed to extract values of the remanent response (Fig. 3d), coercive (Fig. 3e) and nucleation voltages (Fig. 3f). Analysis of this data showed that increasing of the humidity above 60% leads to increasing of the values of coercive and nucleation voltages, while the value of remanent response changes insignificantly. Band excitation resonance frequency stayed constant (about 355 kHz) during the whole loop cycle and for different values of *H*.

Observed parameters of the hysteresis loops are in a good agreement with domain size vs relative humidity data. Interestingly, increase of the effective value of nucleation field is concurrent with the decrease of the domain size.

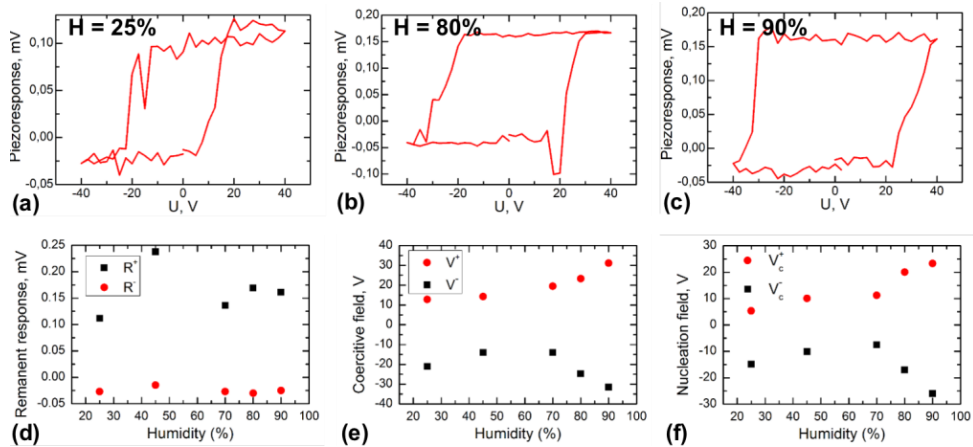

**Figure 3.** (a)-(c) Local hysteresis loops measured at different values of relative humidity. Characteristics of the local hysteresis loops vs humidity: (d) remanent response; (e) coercive and (f) nucleation voltages.

Below, we discuss the observed spectrum of phenomena as affected by the presence and properties of the top water layer. This layer leads to essential redistribution of the electric field produced by biased conductive SPM tip and of course changes switching behavior. Moreover presence of the charge carriers in the top layer creates conditions for enhanced screening of the depolarization field produced by bound charges on the polar surfaces of the sample inside and near the fresh domains. This screening supports growth of the domains at the distances far from the tip. Hence the top water layer significantly changes the switching kinetics due to: delocalization of the external electric field in vicinity of the tip and effective external screening far from the tip.

Below we'll consider first phenomenon – redistribution of the external electric field due to existence of the water at the polar surface in the vicinity of the tip. Meniscus size ranges from tens of nanometers to microns[30] and depending on the relative humidity in the SPM chamber. Unfortunately real geometry of the tip and water meniscus doesn't allow to perform analytical calculations of the electric fields. Hence to calculate the spatial distribution of electric field produced by SPM tip ($E_{tip}$) with water meniscus, we have modeled tip-surface junction in the COMSOL Multiphysics® software package. In the calculations SPM tip has been estimated by cone with rounded apex and radius $R_{tip} = 20$ nm, water meniscus has been estimated by the second order surface with height from the sample surface $h_m$ and length from the tip position $L_m$ (Fig. 4a). In the modeling we assumed $h_m = L_m$. Permittivity of the water and lithium niobate were $\varepsilon_w = 80$ and $\varepsilon_{LN} = 85$ correspondingly.

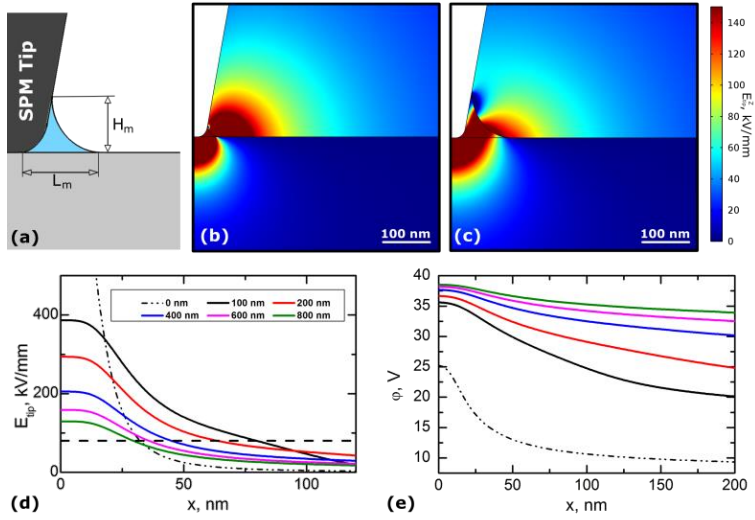

**Figure 4.** Results of COMSOL simulations of the electric field produced by biased SPM tip in presence of water meniscus. (a) Scheme of the model; (b), (c) 2D maps of the spatial distribution of $E_{tip}^Z$ for (b) $h_m = 25$ nm and (c) $h_m = 100$ nm; (d), (e) distributions of $E_{tip}^Z$ and potential φ along polar direction at 10 nm under the surface for different $h_m$.

Computer simulations allowed us to obtain spatial distributions of the z-component of $E_{tip}$ and electric potential φ at depth 10 nm under the sample surface for different values of $h_m$ (Fig. 4). As one can see the appearance of the water meniscus leads to decreasing of the electric field in the area under the tip and it delocalization to long enough distances. To describe obtained simulated data we used well-known point charge model, which has used many times for modeling of electric field produced by biased conductive SPM tip[15,16]. Simulated distributions of $E^Z_{tip}$ for different $h_m$ have been fitted by equation (1), values of the effective tip radius $d^*$ and effective charge $Q^*$ have been extracted as a fitting parameters.

$$E_{tip}^{PC}(r) = \frac{Q^*}{2\pi\varepsilon_0\left(\sqrt{\varepsilon_c\varepsilon_a}+1\right)}\sqrt{\frac{\varepsilon_a}{\varepsilon_c}}\frac{d^*}{\left((d^*)^2+r^2\right)^{3/2}} \quad (1)$$

Dependences of the $Q^*$ and $d^*$ on the meniscus size have demonstrated unexpected view (Fig. 5a). Effective tip radius $d^*$ changes for the small meniscus sizes only (25 – 100 nm) and approaches value of 50 nm for larger sizes. $Q^*$ has short part of the growth at $h_m < 100$ nm and then asymptotically decreases with $h_m$ increase up to the value about $0.5 \times 10^{-15}$ C.

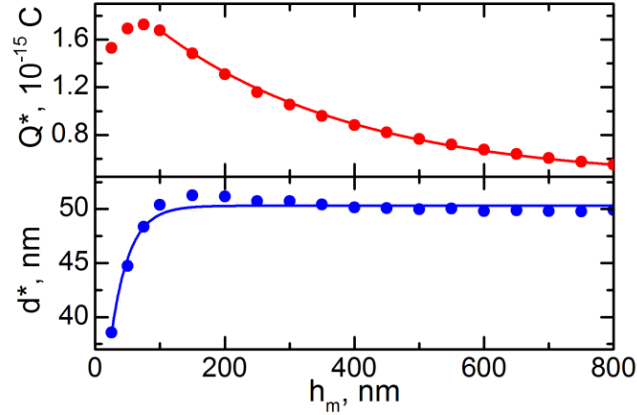

**Figure 5.** Values of the effective charge $Q^*$ and effective tip radius $d^*$ vs. meniscus height in the point charge model.

Extracted values of the effective tip radius and charge allows to describe the growth of the isolated domain with presence of the water meniscus using the model [19]. Actually, domain radius as a function of relative humidity $H$ and bias $U_{sw}$ can be obtained by simple modification the expression derived in Ref[19] allowing for the water meniscus presence:

$$r(U_{sw}, H) = r_0 + \beta d^*(H)\sqrt{\left(\frac{U(U_{sw}, H)}{U_{cr}}\right)^\alpha - 1}, \qquad (2)$$

where α is a power factor $2/3 < \alpha < 2$, dimensionless parametr $\beta$ reflects the tip form-factor and has the order of unity, $U_{cr}$ has sense of a critical voltage for a "dry" ferroelectric surface without water meniscus, ($H = 0$), but it is not so for humidity above 40%, because the water meniscus appearance leads to the partial screening of applied electric voltage $U_{sw}$, and so $U(U_{sw}, H) < U_{sw}$. The "acting" voltage $U(U_{sw}, H)$ contains a "damping" factor. In other words the expression can be rewritten via effective tip parameters $d^*$ and $Q^*$ as

$$U(U_{sw}, H) = \frac{Q^*(U_{sw}, H)}{Ad^*(H)},$$ where $A$ is a factor reflecting the tip geometry.

To fit the simulated data we used trial functions (3) and (4),

$$Q^*(U_{sw}, h_m) = AU_{sw}\left[\chi d_\infty + (d_0 - \chi d_\infty)\exp\left(-\frac{h_m}{h_q}\right)\right] \quad (3)$$

$$d^*(h_m) = d_\infty\left[1 - \frac{d_\infty - d_0}{d_\infty}\exp\left(-\frac{h_m}{h_d}\right)\right] \quad (4)$$

which satisfy the boundary conditions $Q^*(0) = \alpha U_{sw}d_0$, $Q^*(\infty) = \alpha U_{sw}\chi d_\infty$, $d^*(0) = d_0$; $d^*(\infty) = d_\infty$. Here we assume two boundary cases. 1). $h_m = 0$ corresponds to the absence of the water. In this case tip has effective radius $d_0$. Effective tip charge $Q^* = AU_{sw}d_0$ is proportional to bias and tip radius via constant A. 2). $h_m = \infty$ corresponds to the tip under the water (liquid PFM). In this case the tip has effective radius $d_\infty$ which is higher then the tip radius in dry air due to delocalization of the field. Effective charge is still proportional to the tip radius $d_\infty$ and bias $U_{sw}$, but reduced by factor $\chi$ due to higher permittivity of the water $\varepsilon_w = 80$. $h_q$ and $h_d$ are characteristic distances for the effective tip radius and effective charge vs. meniscus size dependences correspondingly.

Fitting of the data (Fig. 6a) gave following values of the constants in the trial functions: $d_0 = 21$ nm; $d_\infty = 50$ nm; $h_d = 28$ nm; $h_q = 282$ nm; $A = 2.6\times10^{-3}$; $\chi = 0.089$.

Obtained value of the $d_0$ is close to $R_{tip}$ used in the COMSOL model, this fact confirms accuracy of the COMSOL calculations. Adding of the water meniscus with size slightly above $R_{tip}$ leads to increasing of effective radius up to 50 nm ($h_d = 28$ nm). While changing of the effective charge $Q^*$ is observed at much longer distances ($h_q \approx 300$ nm). This fact can be attributed to influence of the conical part of the tip, which μm-size is much larger than tip apex (Fig. 4a).

However Eqs. (2)-(4) still can't be used for fitting of the experimental data presented on figure 2b-c, because there isn't a relation between meniscus height and relative humidity in the SPM chamber $h_m(H)$. Unfortunately it can't be directly measured in the used technical configuration and there isn't any publication which contains such investigations on the surface on lithium niobate. So in our calculations we followed results on surface of SiN presented in [30]. We postulated following dependence of the meniscus size on the value of relative humidity:

$$h_m(H) = h_0 \exp\left(-\frac{H_{cr}}{H}\right), \qquad (5)$$

where parameters $h_0$ and $H_{cr}$ are fitting constants (from fitting of the experimental data $h_0 = 2 \times 10^4$ nm; $H_{cr} = 325$).

As a result of Eq.(5) substitution into Eqs.(3)-(4), Eq. (2) acquires the form:

$$r(U_{sw}, h_m) = r_0 + \beta d^*(h_m)\sqrt{\left(\frac{Q(U_{sw}, h_m)}{Ad^*(h_m)U_{cr}}\right)^\alpha - 1} \qquad (6)$$

Here additional fitting parameters are $r_0$, $U_{cr}$ and $\beta$. Value of the constant $\alpha = 2$ is from the linear dependence of the domain radius vs. applied voltage. Fitting of the experimental dependences $r(U_{sw})$ and $r(H)$ by eq. (6) with using (5) showed good agreement between experiment and proposed model (Fig. 5b,c) and gave following values of the of the fitting parameters: $U_{cr} = 3.1$ V; $r_0 = 8$ nm; $\beta = 0.29$. Values of the parameters $U_{cr}$ and $r_0$ predict the formation of the minimal stable domain with radius about 8 nm in dry atmosphere. However minimal experimentally observed domain had size about 35 nm and was observed after application of 20 V bias. Formation of the domains wasn't observed after application of $U_{sw} < 20V$. This fact can be explained by a spontaneous backswitching that led to complete disappearing of the small domains. This phenomenon is often observed during tip-induced switching and leads to formation of the ring-shaped domains[24,25]. Also we should note, constants $A$ and $\beta$ have considerably different values ($A = 2.6 \times 10^{-3}$; $\beta = 0.29$). In spite of the fact they both are proportional by tip geometry, they have different physical meaning $A$ relates tip potential and charge, while $\beta$ – formed domain radius and applied bias.

In addition we should note that we have considered the influence of the top water layer on polarization reversal due to redistribution of the electric field produced by the tip only. But the influence doesn't limited by this phenomenon. In addition, the presence of the top water layer changes all screening conditions by redistribution of the charge carriers across the layer. This phenomenon can be used for explanation of the inconsistence between current results and in refs [24,25]. Here we have studied the congruent LiNbO$_3$ crystal with extremely high value of the coercive field about 21 kV/mm that is 3 times higher than coercive field for Mg doped lithium niobate studied in [24] and more than 6 times higher than coercive filed for stoichiometric lithium niobate studied in [25]. In the both papers growth of the micron-sized domains was observed. At such distances redistribution of the electric filed caused by water meniscus is not so pronounced and can't lead to essential change of the domain kinetics. However external screening caused by charge carriers in the adsorbed surface layer supports the switching far from the tip and leads to formation of the large domains.

In conclusion the local polarization reversal by electric field produced by conductive SPM tip as a function of the relative humidity in SPM chamber has been studied in the single crystal of congruent lithium niobate. Reduction of the formed isolated domain size has been revealed at high values of relative humidity ($H > 50\%$). Local hysteresis measurements reveal appropriate increasing of the coercive and nucleation fields. The observed phenomena have been attributed to existance of the water meniscus in the vicinity of tip − surface contact. Analytical calculations and computer simulations confirm the proposed model and describe process of the domain growth.

## Acknowledgements

A part of this research (A.V.I., S.V.K.) was conducted at the Center for Nanophase Materials Sciences, which is sponsored at Oak Ridge National Laboratory by the Scientific User Facilities Division, Office of Basic Energy Sciences, US Department of Energy. V.Y.S. and A.V.I. acknowledge CNMS user proposal, RFBR and Government of Sverdlovsk region (Grant 13-02-96041-r-Ural) and RFBR (Grant 13-02-01391-a). A.N.M acknowledges the support via bilateral SFFR-NSF project (US National Science Foundation under NSF-DMR-1210588 and State Fund of Fundamental Research of Ukraine, grant UU48/002).